\documentstyle[11pt]{article}
\newlength{\bredde}
\def\slash#1{\settowidth{\bredde}{$#1$}\ifmmode\,\raisebox{.15ex}{/}
\hspace*{-\bredde} #1\else$\,\raisebox{.15ex}{/}\hspace*{-\bredde} #1$\fi}
\newcommand{\beq}{\begin{equation}}
\newcommand{\eeq}{\end{equation}}
\newcommand{\ba}{\begin{array}{ccc}}
\newcommand{\ea}{\end{array}}

\newcommand{\noi}{\vspace{12pt}\noindent}
\newcommand{\lG}{\raise.3ex\hbox{$\stackrel{\leftarrow}{G}$}}
\newcommand{\lU}{\raise.3ex\hbox{$\stackrel{\leftarrow}{U}$}}
\newcommand{\lP}{\raise.3ex\hbox{$\stackrel{\leftarrow}{{\cal P}}$}}
\newcommand{\leta}{\raise.3ex\hbox{$\stackrel{\leftarrow}{\eta}$}}
\newcommand{\lOmega}{\raise.3ex\hbox{$\stackrel{\leftarrow}{\Omega}$}}
\newcommand{\ldr}{\raise.3ex\hbox{$\stackrel{\leftarrow}{\delta^r}$}}

\def\m2{{\mathcal{M}}^{\dagger}{\mathcal{M}}}
\def\mb2{M^2}

\def\beqn{\begin{eqnarray}}
\def\eeqn{\end{eqnarray}}

\def\gtwid{\raise.3ex\hbox{$>$\kern-.75em\lower1ex\hbox{$\sim$}}}
\def\ltwid{\raise.3ex\hbox{$<$\kern-.75em\lower1ex\hbox{$\sim$}}}

\begin{document}
\topmargin -1.4cm
\oddsidemargin -0.8cm
\evensidemargin -0.8cm
\title{\Large{{\bf Partially Quenched Chiral Perturbation Theory \\
and the Replica Method}}}

\vspace{1.5cm}

\author{~\\{\sc P. H. Damgaard} and {\sc K. Splittorff}\\~\\
The Niels Bohr Institute\\ Blegdamsvej 17\\ DK-2100 Copenhagen {\O}\\
Denmark}
\date{\today} 
\maketitle
\vfill
\begin{abstract} 
We describe a novel framework for partially quenched chiral perturbation
theory based on the replica method. The computational rules are
exceedingly simple. We illustrate these rules by
computing the partially quenched chiral condensate to one-loop order.
By considering arbitrary chiral $k$-point functions we show explicitly
to one-loop order the equivalence between this method and the one 
based on supersymmetry. It is possible to go 
smoothly from the conventional replica method to a supersymmetric variant 
by choosing the number of valence quarks to be negative.
\end{abstract}
\vfill

\begin{flushleft}
NBI-HE-00-13 \\
hep-lat/0003017
\end{flushleft}
\thispagestyle{empty}
\newpage

\section{Introduction}

\noi
The question of non-perturbative analytical predictions for quenched
or partially quenched lattice gauge theory computations has been 
thoroughly studied in the context of effective chiral Lagrangians
\cite{BG,BG1,S,CP}. So far the most systematic framework has been the
supersymmetric formulation of Bernard and Golterman \cite{BG,BG1}, which
builds on an idea first introduced in the context of staggered lattice
fermions \cite{M}. Here one introduces $k$ additional
quark species (of conventional statistics) on top of the $N_f$ physical ``sea''
quarks, and $k$ ``ghost'' quarks of opposite statistics to cancel the
effects of the additional quarks. When $N_f$ is taken to vanish this
gives the fully quenched theory, while for $N_f$ non-zero it gives the
partially quenched theory. Both are accessible to a study by Monte Carlo
techniques in lattice gauge theory. The chiral flavor symmetry group is in
that formulation extended to a super Lie group which in perturbation
theory can be taken as SU($N_f+k|k$). (For this reason it is commonly known as
the supersymmetric method although it, as applied, has nothing to do with
space-time supersymmetry, but rather is a graded symmetry.) Based on the
usual assumption of spontaneous chiral symmetry breaking (here extended to
the super group case) the
effective low-energy theory of the lowest-lying hadronic excitations is that
of a chiral Lagrangian, now with fields living on the coset of 
super Lie groups. This
effective Lagrangian can be studied by the conventional methods of chiral
perturbation
theory. In what follows we denote fully and partially quenched chiral 
perturbation theory by QChPT and PQChPT, respectively.

\noi
The supersymmetric framework has also proven to be an efficient means of 
deriving analytical results for the soft part of the Dirac operator
spectrum in finite volume, by taking an appropriate discontinuity of
the partially quenched chiral condensate \cite{OTV,DOTV,TV}. This has brought 
earlier results derived entirely from universal Random Matrix Theory 
\cite{SV} (for a very recent comprehensive review, see ref. 
\cite{VW}) in direct contact with the effective Lagrangian of QCD. 
In particular, a
series of very compact relations that described general $k$-point
spectral correlation functions of low-lying Dirac operator eigenvalues
in terms of effective partition functions with additional quark
species \cite{AD} can now be understood as due to the cancelling pairs
of fermionic and bosonic valence quarks. When taking the same discontinuity
near the origin in PQChPT it has also been shown that one recovers
among other terms the
analytical prediction for the slope of the spectral density of the Dirac
operator at the origin \cite{OTV,DOTV}, a formula first derived in the
QCD case by Smilga and Stern \cite{SS}. The same analysis has 
recently been extended to the two
other major chiral symmetry breaking classes by Toublan and Verbaarschot 
\cite{TV}. There are thus also plenty of
{\em physical} applications of PQChPT that have nothing to
do with the artifacts of the quenched approximation at all.    

\noi
While the supersymmetric approach to QChPT and PQChPT has been well
tested, and is by now quite well understood, it is still of interest to find 
alternative means of formulating the same problem. In particular, 
the supersymmetry itself is not fundamental and not an inherent property of 
QChPT and PQChPT. Indeed, it has recently been shown in the context of
the finite-volume effective chiral Lagrangian related to the soft part
of the Dirac operator spectrum \cite{SV} that the so-called replica method can
provide a useful alternative technique \cite{DS}. Here full or partial
quenching is instead achieved by adding $N_v$ valence quarks 
(of usual statistics),
and then taking the limit $N_v \to 0$ at the end of the calculation. In
ordinary QCD perturbation theory this procedure trivially kills all 
valence quark
loops. In the framework of the effective Lagrangian of Goldstone bosons
it is far from obvious that such a procedure can be carried out explicitly. 
It entails an extension of the chiral symmetry group U($N$) to non-integer 
$N$, and integrals over such a group are not known in
closed form. Nevertheless, it turns out that in series expansions the
required analytical continuation can be carried out explicitly \cite{DS},
and results agree with what was earlier established by the supersymmetric
method \cite{OTV,DOTV}. 
This suggests that also conventional QChPT and PQChPT can be performed
by simply taking the limit $N_v \to 0$. In this paper we shall show
that this is indeed the case. We shall give the very simple Feynman rules,
and explain the intimate relationship to QChPT and PQChPT in the supersymmetric
formulation. As a simple illustration we show how to derive the partially
quenched chiral condensate to one-loop order using this replica method.
This fully or partially quenched chiral condensate is a particularly
convenient observable on which to test the non-perturbative finite-volume 
scaling results discussed above \cite{VMC,DEHN}. The way partially quenched 
chiral perturbation theory smoothly matches on to this regime has been  
explained in ref. \cite{OTV}.

\noi
After providing the Feynman rules, it becomes quite obvious how the replica
method in perturbation theory is equivalent to the 
supersymmetric method. We illustrate a few of the counting rules by
considering a chiral $k$-point function below. 
Mainly out of curiosity, we also show how a variant of the replica
method that is supersymmetric can be used to provide identical results.
This supersymmetric variant is however slightly more cumbersome than
the conventional $N_v\to 0$ replica method, and we do not propose to use that
particular variant for practical calculations.

\section{The replica method}

\noi
As explained above, with the replica method one adds $N_v$ valence quarks
to the QCD Lagrangian, which here can be taken as any SU($N_c\geq 3$) gauge 
theory with $N_f$ physical (sea) quark flavors. Depending on the
applications, it can be convenient to introduce $k$ sets of such valence 
quarks with $k$ different masses $m_{v_j}$, each set containing $N_v$
new quark flavors. The physical quark masses are denoted by $m_f$. 
The QCD partition function with these $kN_v$ additional quark species 
reads
\beq
{\cal Z}^{(N_{f}+kN_{v})} ~=~ 
\int\! [dA]
~\prod_{j=1}^{k}{\det}(i\slash{D} - m_{v_j})^{N_{v}}\prod_{f=1}^{N_{f}}
{\det}(i\slash{D} - m_f) ~e^{-S_{\rm YM}[A]} ~. \label{Zoriginal}
\eeq
This partition function can be viewed as an unnormalized average of $k$
sets of $N_v$ 
identical replicas of the following partition functions of quarks in a
fixed gauge field background $A_{\mu}$:
\beq
{\cal Z}_{v_j} ~\equiv~ \int\! [d\bar{\psi}_jd\psi_j]~ \exp\left[\int\! d^4 x
  \  \bar{\psi}_j(i\slash{D} - m_{v_j})\psi_j\right] 
\eeq
in the sense that
\beq
{\cal Z}^{(N_{f}+kN_{v})} ~=~ 
\int\! [dA]\prod_{j=1}^k\left[{\cal Z}_{v_j}\right]^{N_v} 
\prod_{f=1}^{N_{f}}{\det}(i\slash{D} - m_f) e^{-S_{\rm YM}} 
~.
\eeq
Clearly, if we set $N_v=0$ this just reproduces the original QCD partition 
function. But the theory extended with $kN_v$ additional quark species in
this way is a
generating functional for partially quenched averages of $\bar{\psi}_j\psi_j$
and mixed averages also involving physical quark fields.
One simply sets $N_v$ to zero {\em after} having performed the functional
differentiations
\beq
\chi(m_{v_1},\ldots,m_{v_k},m_{f_1},\ldots,m_{f_l},\{m_f\}) ~\equiv~
\lim_{N_{v}\to 0}
\frac{1}{N^k_{v}}\frac{1}{N^l_{f}}\frac{\partial}{\partial
m_{v_1}}\cdots\frac{\partial}{\partial m_{v_k}}\frac{\partial}
{\partial m_{f_1}}\cdots\frac{\partial}{\partial m_{f_l}}
\ln {\cal Z}^{(N_{f}+N_{v})} ~. \label{generaldef}
\eeq
Technically, it can be convenient to add local sources
for both scalar and pseudoscalar quark bilinears $\bar{\psi}_j(x)\psi_j(x)$
and $\bar{\psi}_j(x)\gamma_5\psi_j(x)$ and similarly for the vector and
axial vector currents (for simplicity taken flavor diagonal).  
If needed, one can of course introduce corresponding sources in the
physical quark sector. Because such terms have no bearing on our
arguments presented below, we shall for simplicity omit them here.

\subsection{Adapting the replica method to the chiral Lagrangian}

\noi
For $N_v$ integer, and $N_f+kN_v$ small enough, chiral symmetry is assumed
to be spontaneously broken according to the standard pattern of
SU$_L(N_f+kN_v)\times$SU$_R(N_f+kN_v) \to$ SU($N_f+kN_v$). The effective
low-energy theory can therefore be described in the entirely
conventional framework of a chiral Lagrangian based on SU($N_f+kN_v$),
with no new assumptions about the pattern of chiral symmetry 
breaking.\footnote{The reader might worry about the assumption that 
$N_f+kN_v$ should be taken small enough for the theory to support 
spontaneous chiral symmetry breaking. Actually, there will be no new 
constraint from this. We simply analyze the chiral Lagrangian
for arbitrary $N_f+kN_v$ even though this Lagrangian is only the low-energy
theory of QCD for $N_f+kN_v$ sufficiently small. However, we take the limit
$N_v\to 0$ in the end. Then we must meet only the usual constraint
that the number of {\em physical} light quarks $N_f$ should be small enough
to lead to spontaneous chiral symmetry breaking.} The cases $N_f=0$
and
$N_f=1$ are obviously very special here. For $N_f=1$ there is not any
spontaneous breaking of chiral symmetry in the theory after taking $N_v$ 
to zero, and the case $N_f=0$ (which would correspond to full quenching)
is so unusual that we shall discuss it separately.

\noi
Having in mind a possibly non-trivial r\^{o}le played by the flavor
singlet meson, the lowest-order effective chiral Lagrangian is
taken to be the usual ${\cal O}(p^2)$ expression
\beq
{\cal L} ~=~ \frac{F^2}{4}{\mbox{\rm Tr}}(\partial_{\mu}U
\partial^{\mu}U^{\dagger}) - 
\frac{\Sigma}{2}{\mbox{\rm Tr}}{\cal M}(U + U^{\dagger})
+ \frac{\mu^2}{2N_{c}}\Phi_0^2 + \frac{\alpha}{2N_{c}}\partial_{\mu}\Phi_0
\partial^{\mu}\Phi_0 ~.
\label{Lchpt}
\eeq
Here the field $U \equiv \exp[i\sqrt{2}\Phi/F]$ is an element of 
SU($N_f+kN_v$), and
we have kept the flavor-singlet field $\Phi_0 \equiv {\mbox{\rm Tr}}\Phi$.
As in the supersymmetric method \cite{BG}, it proves convenient to work
in a ``quark basis'' where $\Phi_{ij}$ corresponds to $\bar{\psi}_i\psi_j$.
With all external sources set to zero, 
this gives a simple propagator for the ``off-diagonal'' mesons corresponding
to $\Phi_{ij} \sim \bar{\psi}_i\psi_j, ~i \neq j$:
\beq
D_{ij}(p^2) ~=~ \frac{1}{p^2+M_{ij}^2} ~,
\eeq
while for the ``diagonal'' mesons $\Phi_{ii} \sim \bar{\psi}_i\psi_i$
the propagator can be written in the form \cite{BG1}
\beq
G_{ij}(p^2) ~=~ \frac{\delta_{ij}}{(p^2+M_{ii}^2)}-\frac{(\mu^2+\alpha
  p^2)/N_c}{(p^2+M_{ii}^2)(p^2+M_{jj}^2){\cal F}(p^2)} ~. 
\label{Gij}
\eeq
Here $M^2_{ij}\equiv (m_i+m_j)\Sigma / F^2$ and
\beq
{\cal F}(p^2)~\equiv~1+\frac{\mu^2+\alpha p^2}{N_c}\left(\sum_{j=1}^k
  \frac{N_v}{p^2+M_{v_jv_j}^2}+\sum_{f=1}^{N_f} 
\frac{1}{p^2+M_{ff}^2}\right) ~.
\label{F}
\eeq
Note that $N_v$ enters as a parameter due to the mass degeneracy of the
valence quarks in each of the $k$ sets. This is exactly what is required in
order to apply the replica method. We remark that the unusual form of
the propagator (\ref{Gij}) just stems from using the quark basis and
including the flavor singlet field $\Phi_0 = {\mbox{\rm Tr}}\Phi$, and not 
from any peculiarities of partial quenching. Although we borrow the result
(\ref{Gij}) from ref. \cite{BG1}, it is also unrelated to the supersymmetry 
of the method discussed there. 

\noi
By including the $\Phi_0$ field in the Lagrangian we have  kept open the
possibility of studing various expansion schemes (see $e.g.$ the second
reference of \cite{BG}). The $\Phi_0$ terms affect only $G_{ij}$. For
$G_{ii}$ the flavor-singlet $\Phi_0$ can give rise to double  
poles in the partially quenched limit, but the appearance of such double poles
is not special to the replica 
method. Indeed such double poles are also present in the supersymmetric
formulation where a thorough study has been done \cite{BG,BG1,S}.  
As we prove in the next section the two 
formulations have equivalent perturbative expansions. The appearance 
of the double pole in the replica method is therefore completely analogous to
the case of the supersymmetric formulation. In particular, we note that
also in the replica formalism the case $N_f=0$ is quite special since in
that case ${\cal F}(p^2)$ simply becomes unity, and the double pole in
$G_{ii}$ is unavoidable. Moreover, in just that case there is no decoupling 
as the scale $\mu$ is sent to infinity.

\noi
In Table \ref{tabProp} we give the explicit relation
between the Feynman rules based on the replica method, and those based
on the supersymmetric formulation. The supersymmetry Feynman rules are
supplemented by the standard relative 
minus sign between boson and fermion loops. Despite
the additional minus signs in the Feynmann rules of the supersymmetric 
formulation, the Green functions are identical in the two formulations. 
As we show below, the signs due to {\em combinatorics} in the replica 
method match those arising from statistics and the supertrace in the
supersymmetric formulation.

\begin{table}[tb]
\renewcommand{\arraystretch}{3.5}
\begin{center}
\begin{tabular}{|r|r|r|}
\hline
{\large Propagator} & {\large replica PQChPT} & {\large supersymmetric
PQChPT}\cr
\hline\hline
$D_{ij}(p^2)$   &
$\frac{1}{p^2+M_{ij}^2}$  & $\frac{\epsilon_i}{p^2+M_{ij}^2}$  \\ \hline
$G_{ij}(p^2)$   &
$\frac{\delta_{ij}}{(p^2+M_{ii}^2)}-\frac{(\mu^2+\alpha 
  p^2)/N_c}{(p^2+M_{ii}^2)(p^2+M_{jj}^2){\cal F}(p^2)}$  &
$\frac{\epsilon_i\delta_{ij}}{(p^2+M_{ii}^2)}-\frac{(\mu^2+\alpha 
  p^2)/N_c}{(p^2+M_{ii}^2)(p^2+M_{jj}^2){\cal F}(p^2)}$ \\ \hline \hline 
${\cal F}(p^2)$ & $1+\frac{\mu^2+\alpha p^2}{N_c}\left(\sum_{j=1}^k
  \frac{N_v}{p^2+M_{v_jv_j}^2}+\sum_{f=1}^{N_f} 
\frac{1}{p^2+M_{ff}^2}\right)$ & $1+\frac{\mu^2+\alpha
p^2}{N_c}\sum_{f=1}^{N_f+2k} 
\frac{\epsilon_i}{p^2+M_{ii}^2}$ \\ \hline
\end{tabular}
\end{center}
\caption{ \label{tabProp}
The propagators for replica PQChPT, SU$(N_f+kN_v)$, and supersymmetric PQChPT,
SU$(N_f+k|k)$. The sign
$\epsilon_i$ is defined as $\epsilon_i\equiv1$ for $i=1,\ldots,N_f+k$ and
$\epsilon_i\equiv-1$ for $i=N_f+k+1,\ldots,N_f+2k$. Note that ${\cal F}$
coincides in the partially quenched limit of the two approaches.} 
\end{table}

\section{The equivalence between replica and supersymmetric PQChPT}

\noi
In this section we formulate the equivalence between the generating
functional of PQChPT in the replica and supersymmetric formulations. 
The equivalence proof is by default restricted to perturbation theory 
(expressed in terms of the Feynman rules), and we can in principle not
make any statements at the non-perturbative level. But this is as it
should be, as our whole
framework in any case is restricted to chiral perturbation theory.
The Lagrangian itself contains an infinitely long string of interactions
that become relevant with increasing loop order, and we shall only demonstrate
the equivalence at the one-loop level. However, seeing how the equivalence
proof proceeds, it is pretty obvious how to generalize this to arbitrarily
high order.

\noi
Our claim is: The generating
functional of replica PQChPT for $N_f+k N_v$ flavors with $k$ sets of $N_v$
mass-degenerate quarks
is in perturbation theory equivalent to the generating    
functional of supersymmetric PQChPT for $N_f+k$ fermionic and 
$k$ bosonic quarks.

\noi
By {\em equivalence} between the SU$(N_f+kN_v)$ and
the SU$(N_f+k|k)$ generating functionals is meant that the chiral
expansions are equivalent order by order. Of course, the respective
limits, $N_v\to0$ and mass degeneracy between the $k$ bosons and $k$ of the
fermions, are to be introduced at the end of the calculations.
While we believe that this statement is true we will as mentioned above 
only address the
equivalence at the one-loop level. At this one-loop level the contributions
from the ${\cal O} (p^4)$ chiral Lagrangian act as counter terms and we can
base the discussion on the Lagrangian of (\ref{Lchpt}).

\noi
Let us first consider the sea sector. (The term sea sector is used when only
sea quark masses are involved in differentiations of the generating
functional.) For this sector both methods are equivalent to SU$(N_f)$
ChPT. In the replica formulation the contributions from the valence quarks at 
one-loop to any of the correlators  
\beq
\chi(m_{f_1},\ldots,m_{f_l},\{m_f\})
~\equiv~\frac{1}{N^k_{f}}\frac{\partial}{\partial m_{f_1}}\cdots
\frac{\partial}{\partial m_{f_k}}
\ln {\cal Z}^{(N_{f}+N_{v})} ~, \label{basicdef}
\eeq
are necessarily proportional to positive powers of $N_v$. Hence the
dependence on the valence quarks vanishes as $N_v\to0$, leaving the sea sector
equivalent to SU$(N_f)$ ChPT.
The analogous statement in supersymmetric
PQChPT was proven in ref. \cite{BG1}. This equivalence was
formulated as three theorems in that reference. At the risk of making
some oversimplifications we state them compactly as follows: 

\noi
{\em I)}~ The sea sector of SU$(N_f+k|k)$ PQChPT
is equivalent to SU$(N_f)$ ChPT. 

\noi
{\em II)}~ The super-$\eta'$ is equivalent to 
the conventional $\eta'$ of SU$(N_f)$ ChPT. 

\noi
{\em III)}~ The double pole of
$G_{ii}$ arise at a given fermionic quark mass if and only if all fermionic
quarks with this mass are paired up by bosonic quarks.

\noi 
In the supersymmetric formalism theorem {\em I}~ is established by
noting that $k$ of the fermionic quarks and the
$k$ bosonic quarks only appear as virtual loops in the sea sector. Since
these 2$k$ quarks are paired up in masses the virtual loops cancel
explicitly. This cancellation is also responsible for establishing theorem 
{\em II} in the supersymmetric formalism,
only now it takes place in the quark loop corrections to the
$\eta'$-propagator. Finally theorem {\em III}  
follows directly from the structure of the last 
term in $G_{ii}$. We emphasize here that the obvious analogs of
both theorems {\em I} and {\em II} are completely trivial in the 
present replica formalism. Theorem {\em III}, when re-stated in the
language of the replica formalism, stipulate under what circumstances
the potential double pole of $G_{ii}$ is cancelled: By inspection this occurs
when $M_{ii}=M_{ff}$ for at at least one physical meson labelled by $ff$.
The proof of theorem {\em III} is then almost identical in the replica and 
supersymmetric formulations. In the phrasing of refs. \cite{BG,BG1,S} the
double poles can only occur at mass scales that are completely quenched.

\noi
In the remaining quark sectors the equivalence is far less trivial. However,
the supersymmetric bosonic Green functions  
equal the fermionic ones up to a well defined sign. 
So we can focus on the sectors involving fermionic valence quarks.

\noi
The equivalence in these sectors is not just of academic interest. As
mentioned in the introduction, differentiations with
respect to valence quark masses may be related to physical quantities.
For instance the partially quenched chiral condensate for the valence quarks, 
\beq
\Sigma(m_v,\{m_f\}) ~\equiv~ \lim_{N_{v}\to 0}
\frac{1}{N_{v}}\frac{\partial}{\partial m_v}
\ln {\cal Z}^{(N_{f}+N_{v})} ~, \label{firstdef}
\eeq
can be used to determine the Dirac spectral density. 
This density is given by the
discontinuity of the partially quenched chiral condensate
across a cut on the imaginary axis \cite{OTV}:
\beq
\rho(\lambda;\{m_f\}) ~=~ \frac{1}{2\pi}
\left .{\rm Disc}\right |_{m_v = i\lambda}\Sigma(m_v,\{m_f\}) 
~=~ \frac{1}{2\pi}\lim_{\epsilon \rightarrow 0}
[\Sigma(i\lambda+\epsilon,\{m_f\}) - 
\Sigma(i\lambda-\epsilon,\{m_f\})] 
~ . \label{spectdisc}
\eeq
(This identification holds when one considers
$\Sigma(m_v,\{m_f\})$ as a function of a {\em real} mass
$m_v$, and then replaces $m_v \to i\lambda \pm\epsilon$.)

\noi
In the valence sector and the mixed sector the equivalence is established in
two steps.
{\em  First}, notice that the
propagator (\ref{Gij}) of replica PQChPT for $N_v=0$ is identical to the one
for the fermionic sector
of the corresponding supersymmetric PQChPT in the limit where each of the boson
masses is paired up with a fermion mass, see Table \ref{tabProp}. (This
equivalence
holds trivially for the off-diagonal quark anti-quark propagators.) {\em 
  Second},
the signs arising from combinatorics in the replica method is exactly matched 
by the opposite signs of boson and fermion loops occurring in the
supersymmetric formulation.

\noi 
In order to see exactly how the signs come to match in the two approaches, we
explicitly give the derivation of the $k$-point function
in the valence sector. The generalization to the mixed sector follows in
complete analogy.

\subsection{ The one-point function in the valence sector}

In this first example we give the contributions to the valence quark mass
dependent chiral condensate defined in (\ref{firstdef}). We show how the
cancellations that occur exactly match those of the supersymmetric
formulation. It turns out that this simple 1-point function
actually is ideally suited for illustrating the equivalence between
the replica method and the supersymmetric method, as all essential
properties of the propagators and of the combinatorics come into play.

\noi
To evaluate the one-point function we need to introduce just one set of
replica fermions. Explicitly performing the differentiation of the generating 
functional, see Eq. (\ref{firstdef}), or alternatively counting the number
of realizations of quark flow diagrams we have to one-loop 
\beq
\frac{\Sigma(m_v,\{m_f\})}{\Sigma} ~=~ \lim_{N_{v}\to 0}
\frac{1}{N_{v}}(N_v-\frac{1}{F^2}\left(N_v\sum_{f=1}^{N_f}\Delta(M_{vf}^2) 
+N_v(N_v-1)\Delta(M_{vv}^2)+N_v \frac{1}{V}\sum_p G_{vv}(p^2)\right))
\eeq
where 
\beq
\Delta(M_{ij}^2) ~ \equiv ~ \frac{1}{V}\sum_p\frac{1}{p^2+M_{ij}^2} ~ \equiv
~ \frac{1}{V}\sum_pD_{ij}(p^2) 
\eeq
is a one-loop integral of the standard diagonal propagator for the 
off-diagonal mesons,
$\Phi_{ij}\sim\bar{\psi}_i\psi_j$, $i\not=j$. (We write everything in
finite-volume notation, having also in mind applications of the kind
discussed in refs. \cite{OTV,DOTV,TV}.) The first term in
$\frac{1}{V}\sum_pG_{vv}(p^2)$ is simply $\Delta(M_{vv}^2)$. For arbitrary
$N_v$ this term is seen to cancel against the term just before $G_{vv}$. In 
the $N_v\to0$ limit we also get rid of the term proportional to $N_v$, leaving
simply
\beq
\frac{\Sigma(m_v,\{m_f\})}{\Sigma} ~=~ 
1-\frac{1}{F^2}\left(\sum_{f=1}^{N_f}\Delta(M_{vf}^2)- \frac{1}{V}\sum_p
\frac{(\mu^2+\alpha
  p^2)/N_c}{(p^2+M_{vv}^2)(p^2+M_{vv}^2){\cal F}(p^2)}\right) ~ .\label{1-pt}
\eeq
This is completely analogous to the result obtained in the supersymmetric
formulation. In that case a similar cancellation takes place between the
first term in $\frac{1}{V}\sum_pG_{vv}(p)$ and the loop of the meson built up
by the fermionic and bosonic valance quark. It is also instructive to trace
the cancellation of valence quark loops. In the supersymmetric formulation
this cancellation occurs because of a matching boson loop, while in the
present formulation it is due to {\em the lack of a replica fermion}. 
Pictorially speaking, this lack of a replica fermion
acts like a boson. 

\subsection{ The $k$-point function in the valence sector}

As for the condensate, the $k$-fold derivative, $k\geq2$, of 
$\ln{\cal Z}^{(N_f+kN_v)}$ with
respect to each of the valence quark masses is related to the spectral
$k$-point function. The evaluation of the $k$-fold derivative is quite simple
but we need to treat  the case $k=2$ separately. The reason is simple:
The product 
\beq
\sum_{j,k=1}^{N_f+2N_v}\Phi_{ij}(x_1)\Phi_{ji}(x_1)\Phi_{lk}(x_2)\Phi_{kl}(x_2) 
~ , ~ ~ ~ ~  i\not=l
\eeq 
occurring in the two point function includes two connected terms, namely
\begin{displaymath}
\Phi_{v_1v_1}(x_1)\Phi_{v_2v_2}(x_2)\Phi_{v_2v_2}(x_2)\Phi_{v_1v_1}(x_1)
\end{displaymath}
and 
\begin{displaymath}
\Phi_{v_1v_2}(x_1)\Phi_{v_2v_1}(x_2)\Phi_{v_1v_2}(x_2)\Phi_{v_2v_1}(x_1) ~ .
\end{displaymath}
For $k>2$ there is no connected analogue of the latter ``crossed 
diagram'', since $k$ of the
indices must be different (we differentiate with respect to different masses).
The 2-point function is thus different from higher $k$-point
functions because meson loops correspond to just quark-antiquark lines.
In terms of the propagators the two-point function is\footnote{This chiral
2-point function 
has been analyzed in the supersymmetric formulation by Osborn, Toublan, and
Verbaarschot (private communication).} 
\beq
\frac{\chi(m_{v_1},m_{v_2},\{m_f\})}{\Sigma^2} ~ = ~  \lim_{N_{v}\to 0}
\frac{1}{N^2_{v}}\frac{1}{F^4}\left(N^2_v\frac{1}{V}\sum_p D_{v_1v_2}(p^2)
D_{v_2v_1}(p^2) 
+N^2_v \frac{1}{V}\sum_p G_{v_1v_2}(p^2)G_{v_2v_1}(p^2)\right) ~ .
\eeq
Whereas for $k>2$ there is no crossed diagram, and we are left with
\beqn
\frac{\chi(m_{v_1},\ldots,m_{v_k},\{m_f\})}{\Sigma^k} & = &  \lim_{N_{v}\to 0}
\frac{1}{N^k_{v}}(-1)^k\frac{1}{F^{2k}}N^k_v \frac{1}{V}\sum_p
  G_{v_1v_2}(p^2)\cdots G_{v_kv_1}(p^2)~ .
\eeqn
We observe that in both cases the $N_v$-dependence is such that the
limit $N_v \to 0$ becomes trivial.
The corresponding expressions in the supersymmetric formalism are
identical. Note that sea fermion and ``ghost''-loops only appear in the
one-point function.

\section{From replicas to supersymmetry}

\noi
Interestingly, in perturbation theory it is possible to use
a peculiar variant of the replica method that is 
supersymmetric.
This is because all $N_v$-dependence in the propagators and vertices is
entirely parametric. We can thus make replicas of an arbitrary real
number of valence quarks. Moreover, partial quenching can be
achieved not only by taking $N_v \to 0$, but also by taking $N_v$ to
any fixed number of quarks $N'_v$, and re-interpreting the 
remaining $N_f+ N'_v$ as physical quarks (of which it just happens
that at least $N'_v$ are degenerate in mass). Because the $N_v$-dependence is
parametric in perturbation theory, we can trivially go one step further
and consider a partially quenched theory of $N_f$ physical fermions as
the limit $N_v \to -\tilde{N}_v$ of a theory based on $N_f + \tilde{N}_v+N_v$
quarks, out of which the $\tilde{N}_v+N_v$ quarks are degenerate in mass 
$\tilde{m}_v=m_v$. This corresponds to considering the effective theory of 
a fundamental partition function that is partially supersymmetric 
(for simplicity considering only one such set of replica quarks):
\beqn 
\left.{\cal Z}^{(N_{f}+\tilde{N}_{v}+N_v)}\right|_{N_{v}= -\tilde{N}_{v}} &=& 
\left. \int\! [dA]
~{\det}(i\slash{D} - m_{v})^{N_{v}}
\prod_{f=1}^{N_{f}+ \tilde{N}_{v}}
{\det}(i\slash{D} - m_f) ~e^{-S_{\rm YM}[A]}\right|_{N_{v}=-\tilde{N}_{v}} \cr
&=& \int\! [dA] ~\frac{{\det}(i\slash{D} - \tilde{m}_{v})^{\tilde{N}_{v}}}
{{\det}(i\slash{D} - m_{v})^{\tilde{N}_{v}}}
\prod_{f=1}^{N_{f}}
{\det}(i\slash{D} - m_f) ~e^{-S_{\rm YM}[A]} ~.\label{superZ}
\eeqn
At this level the partition function is exactly as the starting point of
the supersymmetric method. However, when we consider the effective
partition function in terms of the Goldstone bosons, the working rules
are entirely different. We keep our Feynman rules of Table 1, and just
remember to take the limit $N_v \to -\tilde{N}_v$ in the end. The fact
that this procedure works is of course a direct consequence of the fact
that in perturbation theory we can
get bosons from fermions by letting the number of (degenerate) species
go from positive to negative (also the ``statistics'' sign of closed
fermion loops relative to closed boson loops comes out right in this way).

\noi
It is instructive to see how this supersymmetric variant of the replica
method works in detail. Consider again our prototype of a Green function,
that of the partially quenched chiral condensate. Using the notation of
above, we find 
\beqn
\frac{\Sigma(m_v,\{m_f\})}{\Sigma} & \equiv & \lim_{N_{v}\to -\tilde{N}_{v}
  \atop m_v\to \tilde{m}_v}\frac{\partial}{\partial m_v} {\rm ln} {\cal
Z}^{(N_f+\tilde{N}_v+N_v)}
\nonumber\\ & = & \lim_{N_{v}\to -\tilde{N}_{v} \atop m_v\to \tilde{m}_v}
\frac{1}{N_{v}}(N_v-\frac{1}{F^2}\left(N_v\left[\sum_{f=1}^{N_f}
\Delta(M_{vf}^2) + \tilde{N}_{v}\Delta(M_{v\tilde{v}}^2)\right]\right. 
+N_v(N_v-1)\Delta(M_{vv}^2) \cr
&& \left. + N_v \frac{1}{V}\sum_p G_{vv}(p^2)\right))
\label{superreplica}
\eeqn
where $M^2_{v\tilde{v}} \equiv (m_v+\tilde{m}_v)\Sigma/F^2$, and 
$G_{vv}(p^2)$ is as in Table 1, except for the obvious change that now
\beq
{\cal F}(p^2)~ \equiv~ 1+ \frac{\mu^2+\alpha p^2}{N_c}\left(
  \frac{N_v}{p^2+M_{vv}^2}+ \frac{\tilde{N}_v}{p^2+M_{\tilde{v}\tilde{v}}^2}+
\sum_{f=1}^{N_f} \frac{1}{p^2+M_{ff}^2}\right) ~.
\label{Fsuper}
\eeq
Taking the degenerate mass limit $\tilde{m}_v = m_v$ and letting $N_v
\to - \tilde{N}_v$ we note that terms cancel out exactly as in the previous
$N_v \to 0$ replica method. For instance, in ${\cal F}(p^2)$ the terms
linear in $N_v$ and $\tilde{N}_v$ just cancel each other. In eq. 
(\ref{superreplica}) the term proportional to $N_v^2$, which previously dropped
out trivially in the $N_v \to 0$ limit, is now precisely cancelled by
a similar term proportional to $N_v\tilde{N}_v \to - \tilde{N}_v^2$. All
``unwanted'' terms thus exactly cancel as they should, and we are left
with the correct one-loop result (\ref{1-pt}). As we mentioned earlier,
this example of the one-point function is actually the most instructive
for illustrating the cancellations. The other $k$-point functions clearly 
proceed analogously.

\noi
Although it is thus possible to make a supersymmetric variant of the replica
method, it is obviously rather pointless to do so. The simplest Feynman rules
come from using just the conventional $N_v \to 0$ limit. We also note
that although the starting partition function (\ref{superZ}) is identical
to that forming the basis for the supersymmetric chiral Lagrangian
\cite{BG,BG1}, the effective theory one works with in the analogous
supersymmetric replica scheme is of a very different nature, and has
in fact here only been defined by means of the perturbative expansion.

\section{Conclusions}

\noi
We have shown how
the replica method can been adapted to chiral perturbation theory. This
provides a new and systematic realization of quenched and partially quenched 
chiral perturbation theory. We have demonstrated how the replica method is
equivalent to the supersymmetric formulation in perturbation
theory. This equivalence is quite trivial in the sector of physical quarks,
and has allowed us to extend the three theorems of \cite{BG1} to the 
present replica formulation of PQChPT. The equivalence between the
replica and the supersymmetric formalisms also extends outside the sea
sector. The complete agreement (at least to one-loop order)  
of the two approaches
offers a non-trivial consistency check. In particular, 
the assumed extension of the standard symmetry breaking pattern to the 
supergroup case is avoided in the present context. The fact that
results agree can be taken as independent confirmation of the validity
of both approaches.

\noi
As an equivalent but nevertheless independent formulation of PQChPT the
replica method illustrates the fact that supersymmetry is a technical tool
for quenching rather than of fundamental nature. 
For practical purposes the usefulness of the replica method as compared to
the supersymmetric formulation is perhaps a matter of taste.  The
advantage of having fewer sign-rules using the replica method is to
some extent traded 
for the marginally simpler combinatorics in the supersymmetric formulation.

\noi
Finally, the replica method presented here gives the background and the
justification for the 
rules observed by Colangelo and Pallante in \cite{CP}. Within the
supersymmetric formulation they studied fully
quenched chiral perturbation theory to one loop. 
Based on an explicit calculation of the divergent parts of the 
generating functional for both SU$(k|k)$ (and the additional U(1) of
the $\Phi_0$) and
standard SU$(N_f)$ chiral perturbation theory (without the $\Phi_0$),
they proposed a set of rules for writing down 
 large parts of the SU$(k|k)$ generating functional from that of SU$(N_v)$. 
The equivalence between the SU$(k|k)$  and SU$(N_v \to 0)$ theories (when the
$\Phi_0$ is included in both), is a special case of the general equivalence
established here. This formally establishes the rules suggested
in \cite{CP} and furthermore shows that the terms missing in 
SU$(N_f\to0)$ chiral perturbation are just those produced by including
the $\Phi_0$. The procedure to compute in partially quenched chiral
perturbation theory to any order is now extremely simple. 
One must take a usual chiral SU($N_f+N_v$) chiral Lagrangian and add the
contributions from $\Phi_0$. For example, to order $p^6$ the whole list
of divergent contributions in the case of a degenerate SU($N_v$) theory
is provided in ref. \cite{BCE}. This can form the basis for a fully quenched
calculation once the contributions from the flavor singlet have been
included (for a discussion of the large-$N_c$ limit, see $e.g.$ ref. 
\cite{L}).

\noi
{\sc Acknowledgement:}\\
This work was supported in part by EU TMR grant no. ERBFMRXCT97-0122.

\end{document}